\documentclass[prd,onecolumn,nofootinbib,showkeys]{revtex4}
\usepackage{bm,amsmath,amssymb,graphicx}

\begin{document}

\title{BIG BOUNCE AND CLOSED UNIVERSE FROM SPIN AND TORSION}
\author{{\bf Gabriel Unger}$^{1,2}$}
\altaffiliation{gunge1@seas.upenn.edu}
\author{{\bf Nikodem Pop{\l}awski}$^1$}
\altaffiliation{NPoplawski@newhaven.edu}
\affiliation{$^{1}$Department of Mathematics and Physics, University of New Haven, 300 Boston Post Road, West Haven, CT 06516, USA}
\affiliation{$^{2}$Department of Mechanical Engineering and Applied Mechanics, University of Pennsylvania, 220 South 33rd Street, Philadelphia, PA 19104, USA}

\noindent
{\em The Astrophysical Journal}\\
Vol. {\bf 870}, No. 2, 78 (2019)\\
\copyright\,The American Astronomical Society
\vspace{0.4in}

\begin{abstract}
We analyze the dynamics of a homogeneous and isotropic universe in the Einstein--Cartan theory of gravity.
The coupling between the spin and torsion prevents gravitational singularities and replaces the Big Bang with a nonsingular big bounce, at which the universe transitions from contraction to expansion.
We show that a closed universe exists only when the product of the scale factor and temperature is higher than a particular threshold, contrary to a flat universe and an open universe, which are not restricted.
During inflation, this product must increase to another threshold, so that the universe can reach dark-energy acceleration.
\end{abstract}

\keywords{cosmology: theory, early universe, gravitation.}
\maketitle

\section{Cosmology in Einstein--Cartan (EC) gravity}

The simplest mechanism generating a nonsingular big bounce and inflation, involving only one unknown parameter and no hypothetical fields, arises in the EC theory of gravity \cite{ApJ}.
EC is the simplest and most natural theory of gravity with torsion, with the Lagrangian density for the gravitational field proportional to the Ricci scalar, as in general relativity.
The conservation law for the total (orbital plus spin) angular momentum of fermions in curved spacetime, consistent with the Dirac equation, requires that the antisymmetric part of the affine connection, the torsion tensor \cite{Schr}, is not constrained to zero \cite{req1,req2}.
Instead, torsion is determined by the field equations obtained from varying the action with respect to the torsion tensor \cite{SK1,SK2,SK3,SK4,Lord,EC1,EC2,EC4,EC5,EC6}.
In EC, the spin of fermions is the source of torsion.
The multipole expansion \cite{Pap} of the conservation law for the spin tensor in EC gives a spin tensor that describes fermionic matter as a spin fluid (ideal fluid with spin) \cite{NSH}.
Once the torsion is integrated out, EC reduces to general relativity with an effective spin fluid as a matter source \cite{EC1,EC2,EC4,EC5,EC6}.
The effective energy density and pressure of a spin fluid are given by
\begin{equation}
\tilde{\epsilon}=\epsilon-\alpha n_\textrm{f}^2,\quad\tilde{p}=p-\alpha n_\textrm{f}^2,
\label{eff}
\end{equation}
where $\epsilon$ and $p$ are the thermodynamic energy density and pressure, $n_\textrm{f}$ is the number density of fermions, and $\alpha=\kappa(\hbar c)^2/32$ \cite{HHK}.

The negative corrections from the spin-torsion coupling in (\ref{eff}) generate gravitational repulsion, which prevents the formation of gravitational singularities and replaces the Big Bang with a nonsingular bounce, at which the universe transitions from contraction to expansion \cite{avert1,avert2,avert3}.
These corrections lead to a violation of the strong energy condition by the spin fluid when $\epsilon+3p-4\alpha n_\textrm{f}^2$ drops below 0, thus evading the singularity theorems \cite{HHK}.
Accordingly, this violation could be thought of as the cause of the bounce.
The dynamics of the EC universe filled with a spin fluid (\ref{eff}) has been studied in \cite{spinfluid1,spinfluid2,spinfluid3,spinfluid4,spinfluid5}, with a parity-violating extension in \cite{MZK}, and with torsion coupled to the spinor field in \cite{nonsing}.
The expansion of the closed universe with torsion and quantum particle production shortly after a bounce is almost exponential for a finite period of time, explaining inflation \cite{ApJ}.
Depending on the particle production rate, the universe may undergo several bounces until it produces enough matter to reach a size where the cosmological constant starts cosmic acceleration.
This expansion also predicts the cosmic microwave background radiation parameters that are consistent with the Planck 2015 observations \cite{Planck2015,Planck2016}, as was shown in \cite{SD}.

The avoidance of singularities can also occur in cosmological models based on Riemann--Cartan geometries without spin density: Poincar\'{e} gauge theories with quadratic terms in curvature and torsion \cite{quad1,quad2}, scalar-tensor theories with torsion \cite{st1,st2}, and higher-dimensional geometries with torsion \cite{dim}.
Therefore, it seems to be a generic feature of the Riemann--Cartan spacetime, rather than a particular feature of EC.
We consider EC because it has other interesting consequences.
The spin-torsion coupling modifies the Dirac equation, adding a term that is cubic in spinor fields \cite{HD}.
As a result, fermions must be spatially extended \cite{non1,non2}, which could eliminate infinities arising in Feynman diagrams involving fermion loops.
In the presence of torsion, the four-momentum operator components do not commute and thus the integration in the momentum space in Feynman diagrams must be replaced with the summation over the discrete momentum eigenvalues.
The resulting sums are finite: torsion naturally regularizes ultraviolet-divergent integrals in quantum electrodynamics \cite{toreg1,toreg2}.
Torsion may also explain the matter-antimatter asymmetry and dark matter \cite{anti}, and the cosmological constant \cite{exp}.

The analysis in \cite{ApJ} considered a closed, homogeneous, and isotropic universe in EC.
However, the calculations of the maximum temperature and the minimum scale factor at a bounce neglected the factor $k=1$ in the Friedmann equations (which is justified during and after inflation but not at a bounce before inflation), de facto considering a flat universe.
In this article, we refine these calculations by taking $k$ into account and analyzing the expansion of the universe for all three cases: $k=1$ (closed universe), $k=0$ (flat universe), and $k=-1$ (open universe).
We discover that a closed universe exists only when the product of the scale factor and temperature is higher than a particular threshold, whereas open and flat universes are not restricted by such a condition.
Accordingly, a closed universe forms in a region of space within a trapped null surface \cite{ApJ} when this threshold is reached.

\section{Dynamics of scale factor and temperature}

If we assume that the universe is homogeneous and isotropic, then it is described by the Friedmann--Lema\^{i}tre--Robertson--Walker metric in the isotropic spherical coordinates \cite{LL}:
\begin{equation}
ds^2=c^2 dt^2-\frac{a^2(t)}{(1+kr^2/4)^2}(dr^2+r^2 d\theta^2+r^2\sin^2\theta\,d\phi^2),
\label{metric}
\end{equation}
where $a(t)$ is the scale factor as a function of the cosmic time $t$.
The Einstein field equations for this metric become the Friedmann equations \cite{LL}:
\begin{equation}
\frac{\dot{a}^2}{c^2}+k=\frac{1}{3}\kappa\epsilon a^2
\label{energy}
\end{equation}
and
\begin{equation}
\frac{\dot{a}^2+2a\ddot{a}}{c^2}+k=-\kappa pa^2,
\label{Fri}
\end{equation}
where a dot denotes the derivative with respect to $t$ and $\kappa=8\pi G/c^4$.
Multiplying the first Friedmann equation by $a$ and differentiating over time, and subtracting from it the second Friedmann equation multiplied by $\dot{a}$ gives an equation that has the form of the first law of thermodynamics for an adiabatic universe:
\begin{equation}
\frac{d}{dt}(\epsilon a^3)+p\frac{d}{dt}(a^3)=0.
\end{equation}
For EC, the Friedmann equations have the same form but the energy density and pressure are replaced by $\tilde{\epsilon}$ and $\tilde{p}$ \cite{spinfluid1,spinfluid2,spinfluid4,spinfluid5}:
\begin{equation}
\frac{\dot{a}^2}{c^2}+k=\frac{1}{3}\kappa(\epsilon-\alpha n_\textrm{f}^2)a^2
\end{equation}
and
\begin{equation}
\frac{d}{dt}[(\epsilon-\alpha n_\textrm{f}^2)a^3]+(p-\alpha n_\textrm{f}^2)\frac{d}{dt}(a^3)=0.
\label{cont}
\end{equation}

The spin fluid in the early universe is formed by an ultrarelativistic matter in kinetic equilibrium, for which $\epsilon=h_\star T^4$, $p=\epsilon/3$, and $n_\textrm{f}=h_{n\textrm{f}}T^3$, where $T$ is the temperature of the universe, $h_\star=(\pi^2/30)(g_\textrm{b}+(7/8)g_\textrm{f})k_\textrm{B}^4/(\hbar c)^3$, and $h_{n\textrm{f}}=(\zeta(3)/\pi^2)(3/4)g_\textrm{f}k_\textrm{B}^3/(\hbar c)^3$ \cite{Ric}.
The quantities $g_\textrm{b}$ and $g_\textrm{f}$ are the numbers of spin states for all elementary bosons and fermions, respectively.
For standard-model particles, $g_\textrm{b}=29$ and $g_\textrm{f}=90$.
In the presence of spin and torsion, the first Friedmann equation is therefore \cite{ApJ,tor}
\begin{equation}
\frac{{\dot{a}}^2}{c^2}+k=\frac{1}{3}\kappa(h_\star T^4-\alpha h_{n\textrm{f}}^2 T^6)a^2.
\label{nikoeq1}
\end{equation}
The first law of thermodynamics (\ref{cont}) gives \cite{ApJ}
\begin{equation}
\Bigl(\frac{\dot{a}}{a}+\frac{\dot{T}}{T}\Bigr)\Bigl(1-\frac{3\alpha h_{n\textrm{f}}^2}{2h_\star}T^2\Bigr)=0,
\end{equation}
which yields
\begin{equation}
\frac{\dot{a}}{a}+\frac{\dot{T}}{T}=0.
\label{aT}
\end{equation}
Remarkably, the last equation is the same as that for the relativistic universe without spin and torsion.

\section{Analysis of solutions for a closed universe}
\label{closed}

Let us consider a closed relativistic universe.
We define nondimensional quantities:
\begin{eqnarray}
& & x=\frac{T}{T_\textrm{cr}}, \label{nont} \\
& & y=\frac{a}{a_\textrm{cr}}, \\
& & \tau=\frac{ct}{a_\textrm{cr}},
\end{eqnarray}
where
\begin{equation}
T_\textrm{cr}=\Bigl(\frac{2h_\star}{3\alpha h_{n\textrm{f}}^2}\Bigr)^{1/2}=9.410\times10^{31}\,\textrm{K}
\end{equation}
and
\begin{equation}
a_\textrm{cr}=\frac{9\hbar c}{8\sqrt{2}}\Bigl(\frac{\alpha h_{n\textrm{f}}^4}{h_\star^3}\Bigr)^{1/2}=3.701\times10^{-36}\,\textrm{m}.
\end{equation}
Henceforth, we will use the dot to denote the derivative with respect to the new time coordinate $\tau$. 
Equation (\ref{nikoeq1}) can be written as
\begin{equation}
\dot{y}^2+1=(3x^4-2x^6)y^2.
\label{Gabe1}
\end{equation}
Equation (\ref{aT}) can be integrated to
\begin{equation}
xy=C,
\label{xy}
\end{equation}
where $C$ is a positive constant (because the scale factor and temperature are greater than 0).
Substitution of this relation into Equation (\ref{nikoeq1}) gives
\begin{equation}
\dot{y}^2+1=\frac{3C^4}{y^2} - \frac{2C^6}{y^4}.
\label{Gabe2}
\end{equation}
Since the left-hand side of this equation is positive, $y$ cannot reach zero because the right-hand side of this equation would have to become negative.
Consequently, a cosmological singularity is never produced for any value of $C$.

The big bounce and big crunch of a closed universe are turning points (there is no expansion or contraction at these points), so therefore they are determined by a condition
\begin{equation}
\dot{y}= 0.
\end{equation}
Using this relation, Equation (\ref{Gabe2}) can be resolved for a quadratic, in terms of $y$ and $C$.
The resulting quadratic equation for $y^2$ is 
\begin{equation}
y^4-3y^2 C^4+2C^6 =0.
\label{quadr}
\end{equation}
The solutions of this equation are:
\begin{equation}
y^2_\pm = \frac{3C^4 \pm \sqrt{9C^8-8C^6}}{2}.
\label{Gabe3}
\end{equation}
At the big bounce, $y=y_-$, and at the big crunch, $y=y_+$.
These turning points of a closed universe exist if the expression under the square root in Equation (\ref{Gabe3}) is positive or zero.
In order for that expression to remain positive or zero, $C$ must be greater than or equal to $\sqrt{8/9}$.
Consequently, an inequality
\begin{equation}
C \ge \sqrt{8/9},
\label{thresh}
\end{equation}
equivalent to
\begin{equation}
aT \ge \sqrt{8/9}a_\textrm{cr}T_\textrm{cr},
\label{create}
\end{equation}
is a necessary condition for creating a closed universe in a region of space with the local values of $a$ and $T$.

If $C = \sqrt{8/9}$, then the turning points coincide, $y_-=y_+$, and the universe is stationary (no expansion or contraction) with the constant value of the scale factor of $y=\sqrt{32/27}$.
Since the values of $y$ and $C$ are now both known, the corresponding constant value of $x$ can be found using Equation (\ref{xy}) to be $\sqrt{3}/2$.
Such a stationary universe has a constant scale factor $a=\sqrt{32/27}a_\textrm{cr}$ and temperature $T=(\sqrt{3}/2)T_\textrm{cr}$.

If $C$ is greater than $\sqrt{8/9}$, then the universe has two turning points and both the big bounce and big crunch occur.
The value of $\dot{y}^2$ is nonnegative in the range from $y=y_-$ to $y=y_+$, so therefore the universe oscillates between $y=y_-$ and $y=y_+$.
If $C>\sqrt{8/9}$, then Equation (\ref{Gabe2}) can be rearranged to give 
\begin{equation}
y^2_\pm = 3C^4\Bigl[\frac{1\pm \sqrt{1-\frac{8}{9C^2}}}{2}\Bigr].
\label{tpGabe2}
\end{equation}
In the limit $C\gg 1$, using the formula $(1-x)^n \approx 1+nx$ for $|x|\ll 1$ gives 
\begin{equation}
y^2 = 3C^4\Bigl[\frac{1\pm(1-\frac{4}{9C^2})}{2}\Bigr].
\end{equation}
Accordingly, $y_-\approx \sqrt{2/3}C$ and $y_+\approx \sqrt{3}C^2$.

The absolute minimum value for $y$ among all possible values of $C$ can be determined from a condition $dy^2_-/dC=0$, giving 
\begin{equation}
C = 1.
\end{equation}
For this value of $C$, the nondimensionalized minimum scale factor and the corresponding maximum temperature are
\begin{equation}
x = 1, \quad y_\textrm{min} = 1.
\end{equation}
Accordingly, the constant $a_\textrm{cr}$ is equal to the least possible value of the scale factor of a closed universe in EC.
A closed universe with spin and torsion is nonsingular ($y \ge 1$ and thus $y>0$).

The squared values of $y_\textrm{min}$, $y_\textrm{max}$, $x_\textrm{min}$, and $x_\textrm{max}$ for different values of $C$ are shown in Table~\ref{table1}.
The greatest possible value of the temperature in a closed universe in EC is $\sqrt{3/2}T_\textrm{cr}$.
When $C$ is much greater than $1$, the universe can expand by a factor of $9C^2/2$ and its temperature can decrease by the same factor, reaching the value at which the transition from the radiation (relativistic) domination to the matter (nonrelativistic) domination occurs. 

\begin{table}[ht]
\begin{center}
\begin{small}
\begin{tabular}{ |c|c|c|c|c|}
\hline
$C$  & \quad $y^2_\textrm{min}$ & \quad $y^2_\textrm{max}$ & \quad
$x^2_\textrm{max}$ & \quad
$x^2_\textrm{min}$\\
\hline
$\sqrt{8/9}$ & $\frac{32}{27}$ & $\frac{32}{27}$ & $\frac{3}{4}$ & $\frac{3}{4}$ \\
1 & 1 & 2 & 1 & $\frac{1}{2}$ \\
$\gg1$ & $\frac{2C^2}{3}$ & $3C^4$ & $\frac{3}{2}$ & $\frac{1}{3C^2}$ \\
\hline
\end{tabular}\caption{The minima and maxima of the nondimensionalized temperature $x$ and scale factor $y$ for different values of the integration constant $C$ in a closed universe. The domain of $C$ is $[\sqrt{8/9},\infty)$.}
\label{table1}
\end{small}
\end{center}
\end{table}

\begin{figure}
\centering
\includegraphics[width=0.7\textwidth]{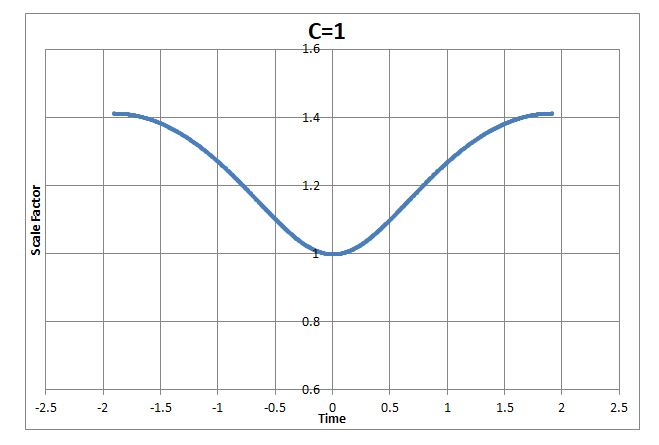}
\caption{Nondimensionalized scale factor $y$ as a function of the nondimensionalized time $\tau$ near a bounce with $C=1$.
The time $\tau=0$ is set at the bounce.}
\label{fig1}
\end{figure}

\begin{figure}
\centering
\includegraphics[width=0.7\textwidth]{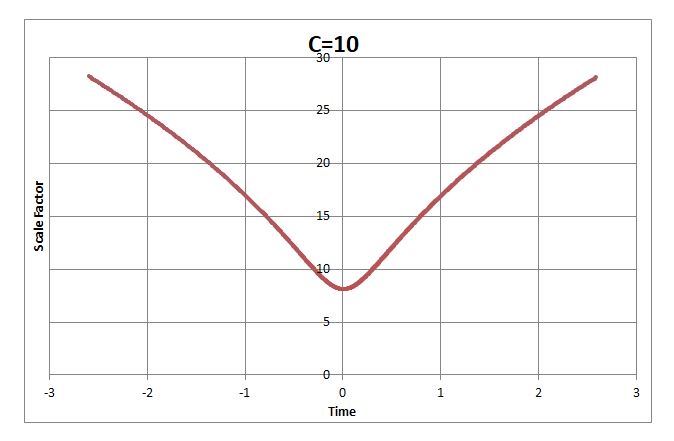}
\caption{Nondimensionalized scale factor $y$ as a function of the nondimensionalized time $\tau$ near a bounce with $C=10$.
The time $\tau=0$ is set at the bounce.}
\label{fig10}
\end{figure}

\begin{figure}
\centering
\includegraphics[width=0.7\textwidth]{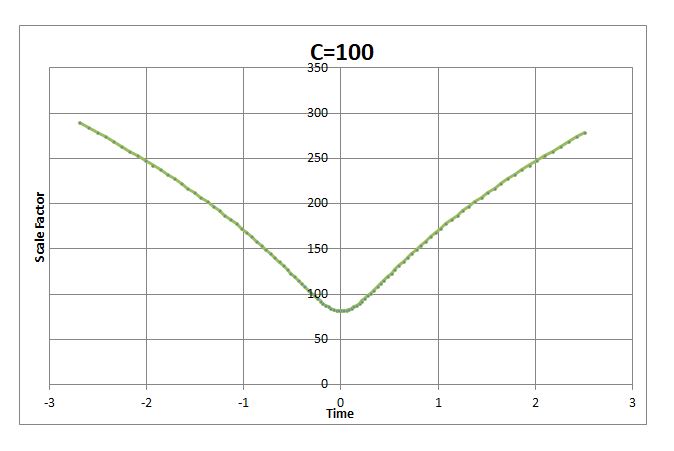}
\caption{Nondimensionalized scale factor $y$ as a function of the nondimensionalized time $\tau$ near a bounce with $C=100$.
The time $\tau=0$ is set at the bounce.}
\label{fig100}
\end{figure}

\begin{figure}
\centering
\includegraphics[width=0.7\textwidth]{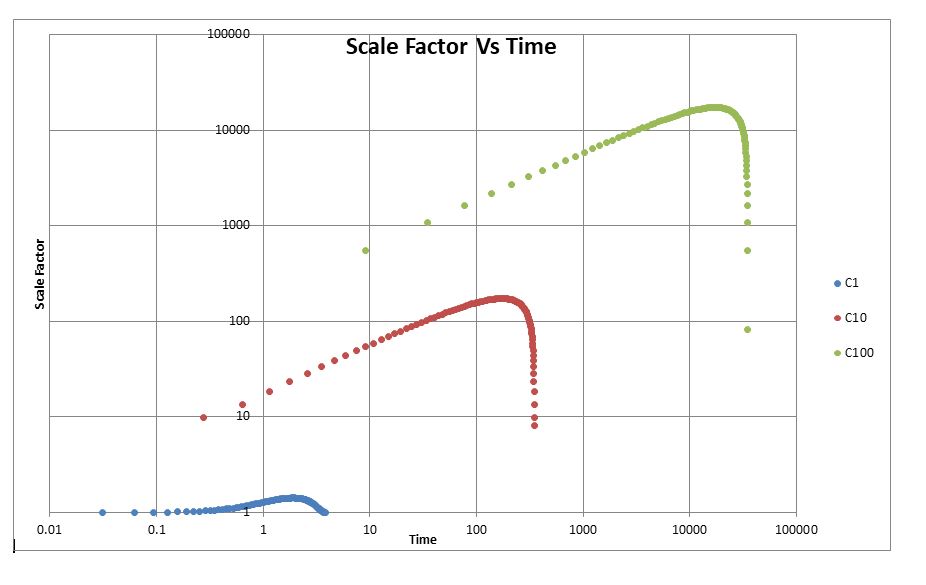}
\caption{Nondimensionalized scale factor $y$ as a function of the nondimensionalized time $\tau$ for one cycle with $C=1,10,100$ using a logarithmic scale.}
\label{figlog}
\end{figure}

Equation (\ref{Gabe2}) can be solved analytically.
We substitute
\begin{equation}
y=[E-F\cos(2\theta)]^{1/2},\quad\dot{y}=[E-F\cos(2\theta)]^{-1/2}F\sin(2\theta)\dot{\theta},
\label{par1}
\end{equation}
where $E$ and $F$ are positive constants such that $E>F$, obtaining
\begin{equation}
[F^2\sin^2(2\theta)\dot{\theta}^2][E-F\cos(2\theta)]+[E-F\cos(2\theta)]^2-3C^4[E-F\cos(2\theta)]+2C^6=0.
\end{equation}
Putting $2E=3C^4$ and $E^2-F^2=2C^6$ reduces this equation into
\begin{equation}
\dot{\theta}^2[E-F\cos(2\theta)]=1,
\end{equation}
which integrates to
\begin{equation}
\tau=\int_0^\theta[E-F\cos(2\theta)]^{1/2}d\theta=(E-F)^{1/2}\int_0^\theta[1+\xi^2\sin^2\theta]^{1/2}d\theta,
\label{par2}
\end{equation}
where $\xi^2=2F/(E-F)$, and $\theta=0$ at $t=0$.
The integral in this equation is the elliptic integral of the second kind with an imaginary $\xi$.
Therefore, the time dependence $y(\tau)$ of the scale factor is given in the parametric form by the first equation in (\ref{par1}) and Equation (\ref{par2}) with
\begin{equation}
E=3C^4/2,\quad F=\sqrt{9C^8/4-2C^6}.
\end{equation}
The parameter $\theta$ runs from 0 (bounce) through $\pi/2$ (crunch) to $\pi$ (next bounce) for one cycle.
The minimum and maximum scale factors are given by $y_\pm=\sqrt{E\pm F}$, in agreement with (\ref{tpGabe2}).
The value of $F$ is real, giving the condition (\ref{thresh}).
Figures \ref{fig1}, \ref{fig10}, and \ref{fig100} show the nondimensionalized scale factor $y$ as a function of the nondimensionalized time $\tau$ near a nonsingular, smooth bounce with $C=1$, $C=10$, and $C=100$, respectively.
Figure \ref{figlog} shows the nondimensionalized scale factor $y$ as a function of the nondimensionalized time $\tau$ for one cycle with $C=1,10,100$ using a logarithmic scale.

The results of this section show that a necessary condition for the creation of an expanding universe in a given volume of space is
\begin{equation}
C>\sqrt{8/9}
\label{ineq}
\end{equation}
for all points in that volume.
For all allowed values of $C$, a closed universe in nonsingular.

\section{Analysis of solutions for a flat and an open universe}

For a flat relativistic universe, Equation (\ref{Gabe2}) becomes
\begin{equation}
\dot{y}^2=\frac{3C^4}{y^2} - \frac{2C^6}{y^4}.
\end{equation}
The corresponding quadratic equation for $y^2$ at the turning points is 
\begin{equation}
3y^2-2C^2 =0.
\end{equation}
The physical solution of this equation is
\begin{equation}
y = \sqrt{2/3}C,
\end{equation}
which is the minimum value of the nondimensionalized scale factor.
This value is equal to the nondimensionalized scale factor at a bounce for the case of a closed universe in the limit $C\gg 1$.
Consequently, a flat universe in EC is also nonsingular, because $C$ is positive.
Since a flat universe has only one turning point (at a bounce), it expands to infinity.
Contrary to a closed universe, there are no further restrictions on the value of $C$.

For an open relativistic universe, Equation (\ref{Gabe2}) becomes
\begin{equation}
    \dot{y}^2-1=\frac{3C^4}{y^2} - \frac{2C^6}{y^4}.
\end{equation}
The corresponding quadratic equation for $y^2$ at the turning points is 
\begin{equation}
    y^4+3y^2 C^4-2C^6 =0.
\end{equation}
The physical solution of this equation is
\begin{equation}
    y^2 = \frac{-3C^4+\sqrt{9C^8+8C^6}}{2},
\end{equation}
with $y>0$.
This solution is the minimum value of the nondimensionalized scale factor.
As for a flat universe, an open universe is also nonsingular and has only one turning point (at a bounce), so it expands to infinity.
There are no further restrictions on the value of $C$.

\section{Analysis with cosmological constant}

A flat and an open universe expand to infinity.
Without further considerations, a closed expanding universe has two turning points (provided that $C>\sqrt{8/9}$) and therefore it reaches the maximum value of the scale factor, after which it contracts.
To avoid the big crunch and the subsequent contraction, another term in the first Friedmann equation is needed that can cause the acceleration of a late universe.
The simplest term is given by a cosmological constant, which enters the first Friedmann equation (for relativistic matter) according to
\begin{equation}
    \dot{y}^2 + 1 = \frac{3C^4}{y^2}- \frac{2C^6}{y^4} + \lambda y^2,
\end{equation}
where $\lambda>0$ is the nondimensionalized cosmological constant:
\begin{equation}
\lambda=\frac{1}{3}\Lambda a_\textrm{cr}^2 = 5.0\times10^{-124}.
\end{equation}
This small value results from the small cosmological constant $\Lambda=1.1\times10^{-52}\,\textrm{m}^{-2}$.

For a late universe, where the values of $y$ are large, the $y^{-4}$ component of this equation can be ignored, reducing this equation to 
\begin{equation}
    \dot{y}^2 + 1 = \frac{3C^4}{y^2} + \lambda y^2.
    \label{Gabe6}
\end{equation}
Putting $\dot{y} = 0$ to find the turning points, the equation can then be rewritten as the quadratic given below:
\begin{equation}
    \lambda z^2 - z + 3C^4 = 0,
\end{equation}
where $z=y^2$.
Solving the quadratic shows that in order for no turning point to exist (in a late universe), $\lambda$ must be greater than $1/(12C^4)$.

This condition can also be written as
\begin{equation}
    C > (12\lambda)^{-1/4}.
\end{equation}
A closed universe expands to infinity if the cosmological constant is sufficiently high.
Such expansion is asymptotically exponential: $\dot{y}^2\approx\lambda y^2$ in a late universe gives $y\sim\exp(\sqrt{\lambda}\tau)$.
According to the results of Section \ref{closed} and the condition (\ref{ineq}), a closed universe forms in a given region of space when the local value of $C$ is equal to $\sqrt{8/9}$.
When $C>\sqrt{8/9}$, the universe expands.
To expand to infinity, the universe must have a mechanism to increase the value of $C$ from $\sqrt{8/9}$ to $(12\lambda)^{-1/4}$.
If $(12\lambda)^{-1/4}<\sqrt{8/9}$, then the universe expands to infinity regardless.
A natural and physical mechanism for the growth of $C$ is provided by quantum particle-pair production in strong gravitational fields \cite{Zel1,Zel2,Zel3,Zel4,Zel5,Zel6,Zel7}.
This mechanism also generates a brief period of exponential expansion of a very early universe \cite{ApJ}, thus naturally deriving inflation \cite{infl1,infl2,infl3}.

This analysis would be valid for a relativistic universe.
However, a relativistic universe transitions from the radiation-dominated era to the matter-dominated era and becomes nonrelativistic.
Equation (\ref{Gabe6}) becomes
\begin{equation}
    \dot{y}^2 + 1 = \frac{B}{y} + \lambda y^2,
    \label{Gabe7}
\end{equation}
where $B$ is a positive constant.
This transition occurs when
\begin{equation}
    \frac{3C^4}{y^2}=\frac{B}{y}
\end{equation}
and at temperature $T_\textrm{eq}=8.8\times10^3$ K.
Using (\ref{nont}) and (\ref{xy}), we obtain
\begin{equation}
    B=3C^3 x_\textrm{eq}^3,
    \label{equality}
\end{equation}
where
\begin{equation}
    x_\textrm{eq}=\frac{T_\textrm{eq}}{T_\textrm{cr}}=9.4\times10^{-29}.
\end{equation}

A nonrelativistic universe expands to infinity if the cosmological constant is sufficiently high.
The turning points of Equation (\ref{Gabe7}) are given by a cubic equation
\begin{equation}
    y^3-\frac{y}{\lambda}+\frac{B}{\lambda}=0.
\end{equation}
This equation has no real positive solutions if \cite{Lord}
\begin{equation}
    B>\frac{2}{3\sqrt{3\lambda}}.
\end{equation}
Consequently, (\ref{equality}) gives the following condition for the absence of turning points in a late universe:
\begin{equation}
    C>\Bigl(\frac{2}{9\sqrt{3\lambda}}\Bigr)^{1/3}\frac{1}{x_\textrm{eq}}=1.9\times10^{48}.
    \label{infin}
\end{equation}
If this condition is satisfied, the universe expands to infinity.

Numerical analysis of generic gravitational collapse in general relativity shows that each spatial point in the interior of a black hole locally evolves toward the singularity as an independent, spatially homogeneous, closed universe \cite{comp1,comp2}.
In EC, this evolution does not reach a singularity, but instead undergoes a nonsingular bounce after which such a universe expands \cite{ApJ}.
Accordingly, if our Universe is closed, its contraction before the big bounce could correspond to gravitational collapse of matter inside a newly formed black hole existing in another universe \cite{Pat1,Pat2,Pat3,Pat4,Pat5,Pat6,Pat7,Pat8,Pat9,Pat10}.
In this scenario, the formation of our Universe corresponds to the moment when the quantity $C$, representing the product of the scale factor and temperature, begins to satisfy the inequality (\ref{ineq}) in a given volume of space in the black hole.
During inflation, this quantity increases because of quantum particle-pair production in strong gravitational fields, and must reach the threshold (\ref{infin}) so that the Universe could start the observed current acceleration.
If this threshold is not reached, the closed universe contracts to another bounce and starts another cycle of expansion.
The last bounce before reaching the threshold can be regarded as the Big Bang of the Universe.

This work was funded by the University Research Scholar program at the University of New Haven.

\end{document}